\documentclass[pdftex,citeautoscript,aps,pra,superscriptaddress,cite,amsmath,amsfonts,amssymb,twocolumn]{revtex4}
\renewcommand{\vec}[1]{\mathbf{#1}} 
\usepackage[pdftex]{graphicx}
\usepackage{multirow}
\usepackage{wasysym}
\usepackage{setspace}
\usepackage{color}

\renewcommand{\Im}{\operatorname{Im}}

\newcommand{\figref}[1]{Fig.~\ref{fig:#1}}
\newcommand{\Figref}[1]{Figure~\ref{fig:#1}}

\newcommand{\eqnumref}[1]{(\ref{eq:#1})}
\renewcommand{\eqref}[1]{Eq.~\eqnumref{#1}}
\newcommand{\eqreftwo}[2]{Eqs.~(\ref{eq:#1}--\ref{eq:#2})}
\newcommand{\Eqreftwo}[2]{Equations~(\ref{eq:#1}--\ref{eq:#2})}
\newcommand{\Eqref}[1]{Equation~\eqnumref{#1}}

\newcommand{\secref}[1]{Sec.~\ref{sec:#1}}

\def\a{s}
\def\b{s}
\newcommand{\add}[1]{\if\a\b{{\color{red} #1}}\else{#1}\fi}
\newcommand{\comm}[1]{\if\a\b{{\color{blue}\{\small \sc #1\}}}\else{}\fi}
\newcommand{\del}[1]{{\if\a\b{{\color{magenta}[[#1]]}}\else{}\fi}}

\begin{document}

\title{Casimir forces in the time domain: I. Theory}

\author{Alejandro~W. Rodriguez}
\affiliation{Department of Physics,
Massachusetts Institute of Technology, Cambridge, MA 02139}
\author{Alexander~P. McCauley}
\affiliation{Department of Physics,
Massachusetts Institute of Technology, Cambridge, MA 02139}
\author{John D. Joannopoulos}
\affiliation{Department of Physics,
Massachusetts Institute of Technology, Cambridge, MA 02139}
\author{Steven G. Johnson}
\affiliation{Department of Mathematics,
Massachusetts Institute of Technology, Cambridge, MA 02139}

\begin{abstract}
  We introduce a method to compute Casimir forces in arbitrary
  geometries and for arbitrary materials based on the
  finite-difference time-domain (FDTD) scheme. The method involves the
  time-evolution of electric and magnetic fields in response to a set
  of current sources, in a modified medium with frequency-independent
  conductivity.  The advantage of this approach is that it allows one
  to exploit existing FDTD software, without modification, to compute
  Casimir forces.  In this manuscript, part~I, we focus on the
  derivation, implementation choices, and essential properties of the
  time-domain algorithm, both considered analytically and illustrated
  in the simplest parallel-plate geometry. Part~II presents results
  for more complex two- and three-dimensional geometries.
\end{abstract}

\maketitle

\section{Introduction}

In recent years, Casimir forces arising from quantum vacuum
fluctuations of the electromagnetic field~\cite{casimir, Lifshitz80,
  milonni} have become the focus of intense theoretical and
experimental effort~\cite{Boyer74, Lamoreaux97, moh1,
  Mohideen02:lateral, hochan1, pnas, Iannuzzi05, maianeto05,
  Brown-Hayes05, Bordag06, Onofrio06, Emig07:ratchet, Munday07,
  Golestanian08, Genet08, Dobrich08, Munday09, Klimchitskaya09}. This
effect has been verified via many experiments~\cite{bordag01,
  milton04, Lamoreaux05, Capasso07:review}, most commonly in simple,
one-dimensional geometries involving parallel plates or approximations
thereof, with some exceptions~\cite{Chan08}. A particular topic of
interest is the geometry and material dependence of the force, a
subject that has only recently begun to be addressed in
experiments~\cite{Chan08} and by promising new theoretical
methods~\cite{emig01, gies03, Gies06:worldline, Rodriguez07:PRA,
  Rahi07, RahiRo07, Emig07, Dalvit08, Kenneth08, Lambrecht08,
  ReidRo09, Pasquali09}.  For example, recent works have shown that it
is possible to find unusual effects arising from many-body
interactions or from systems exhibiting strongly coupled material and
geometric dispersion~\cite{Antezza06, Rodriguez07:PRL, Zaheer07,
  Rodriguez08:PRL, Milton08}. These numerical studies have been mainly
focused in two-dimensional~\cite{emig03_1, Hertzberg05, Rodrigues06,
  Bordag06} or simple three-dimensional constant-cross-section
geometries~\cite{Dalvit06, Rodriguez07:PRL, Emig07} for which
numerical calculations are tractable.

In this manuscript, we present a simple and general method to compute
Casimir forces in arbitrary geometries and for arbitrary materials
that is based on a finite-difference time-domain (FDTD) scheme in
which Maxwell's equations are evolved in time~\cite{Taflove00}. A
time-domain approach offers a number of advantages over previous
methods. First, and foremost, it enables researchers to exploit
powerful free and commercial FDTD software with no modification. The
generality of many available FDTD solvers provides yet another means
to explore the material and geometry dependence of the force,
including calculations involving anisotropic dielectrics~\cite{Rosa08}
and/or three-dimensional problems. Second, this formulation also
offers a fundamentally different viewpoint on Casimir phenomena, and
thus new opportunities for the theoretical and numerical understanding
of the force in complex geometries.

Our time-domain method is based on a standard formulation in which the
Casimir force is expressed as a contour integral of the
frequency-domain stress tensor~\cite{Lifshitz80}. Like most other
methods for Casimir calculations, the stress tensor method typically
involves evaluation at imaginary frequencies, which we show to be
unsuitable for FDTD.  We overcome this difficulty by exploiting a
recently-developed exact equivalence between the system for which we
wish to compute the Casimir force and a transformed problem in which
all material properties are modified to include
dissipation~\cite{RodriguezMc09:PRL}. To illustrate this approach, we
consider a simple choice of contour, corresponding to a conductive
medium, that leads to a simple and efficient time-domain
implementation. Finally, using a free, widely-available FDTD
code~\cite{Farjadpour06}, we compute the force between two
vacuum-separated perfectly-metallic plates, a geometry that is
amenable to analytical calculations and which we use to analyze
various important features of our method. An illustration of the power
and flexibility of this method will be provided in a subsequent
article~\cite{McCauleyRo09}, currently in preparation, in which we
will demonstrate computations of the force in a number of non-trivial
(dispersive, three-dimensional) geometries as well as further
refinements to the method.


\section{Method}

In what follows, we derive a numerical method to compute the Casimir
force on a body using the FDTD method. The basic steps involved in
computing the force are:
\begin{itemize}
  \item[(1)] Map the problem exactly onto a new problem with
    dissipation given by a frequency-independent conductivity
    $\sigma$.
  \item[(2)] Measure the electric $\vec{E}$ and magnetic $\vec{H}$
    fields in response to current pulses placed separately at each
    point along a surface enclosing the body of interest.
  \item[(3)] Integrate these fields in space over the enclosing
    surface and then integrate this result, multiplied by a known
    function $g(-t)$, over time $t$, via~\eqref{time-force}.
\end{itemize}

The result of this process is the exact Casimir force (in the limit of
sufficient computational resolution), expressed via \eqref{time-force}
and requiring only the time-evolution of \eqreftwo{FDTD1}{FDTD2}.

In this section, we describe the mathematical development of our
time-domain computational method, starting from a standard formulation
in which the Casimir force is expressed as a contour integral of the
frequency-domain stress tensor.  We consider the frequency domain for
derivation purposes only, since the final technique outlined above
resides entirely in the time domain.  In this framework, computing the
Casimir force involves the repeated evaluation of the photon Green's
function $G_{ij}$ over a surface $S$ surrounding the object of
interest. Our goal is then to compute $G_{ij}$ via the FDTD
method. The straightforward way to achieve this involves computing the
Fourier transform of the electric field in response to a short pulse.
However, in most methods a crucial step for evaluating the resulting
frequency integral is the passage to imaginary frequencies,
corresponding to imaginary time.  We show that, in the FDTD this,
gives rise to exponentially growing solutions and is therefore
unsuitable. Instead, we describe an alternative formulation of the
problem that exploits a recently proposed equivalence in which contour
deformations in the complex frequency-domain $\omega(\xi)$ correspond
to introducing an effective dispersive, dissipative medium at a real
``frequency'' $\xi$. From this perspective, it becomes simple to
modify the FDTD Maxwell's equations for the purpose of obtaining
well-behaved stress tensor frequency integrands. We illustrate our
approach by considering a contour corresponding to a medium with
frequency-independent conductivity $\sigma$. This contour has the
advantage of being easily implemented in the FDTD, and in fact is
already incorporated in most FDTD solvers. Finally, we show that it is
possible to abandon the frequency domain entirely in favor of
evaluating the force integral directly in the time domain, which
offers several conceptual and numerical advantages.

\subsection{Stress Tensor Formulation}

The Casimir force on a body can be expressed~\cite{Lifshitz80} as an
integral over any closed surface $S$ (enclosing the body) of the mean
electromagnetic stress tensor $\langle
T_{ij}(\vec{r},\omega)\rangle$. Here $\vec{r}$ denotes spatial position
and $\omega$ frequency.  In particular, the force in the $i$th direction
is given by:
\begin{equation}
    F_i = \int_0^\infty d\omega \oiint_{\mathrm{S}} \sum_j \langle
    T_{ij}(\vec{r},\omega) \rangle \, dS_j \,,
\label{eq:Force}
\end{equation}

The stress tensor is expressed in terms of correlation functions of
the the field operators $\langle
E_i(\vec{r},\omega)E_j(\vec{r}^\prime,\omega)\rangle$ and $\langle
H_i(\vec{r},\omega)H_j(\vec{r}^\prime,\omega)\rangle$:
\begin{multline}
\left\langle T_{ij} (\vec{r},\omega) \right\rangle = \\
\mu(\vec{r},\omega) \left[ \left\langle
H_{i}(\vec{r})\,H_{j}(\vec{r})\right\rangle_\omega
-\frac{1}{2}\delta_{ij}\sum_k\left\langle
H_{k}(\vec{r})\,H_{k}(\vec{r})\right\rangle_\omega\right] \\ +
\varepsilon(\vec{r},\omega) \Big[ \left\langle
  E_{i}(\vec{r})\,E_{j}(\vec{r})\right\rangle_\omega
  -\frac{1}{2}\delta_{ij}\sum_k \left\langle
  E_k(\vec{r})\,E_k(\vec{r})\right\rangle_\omega \Big] \,,
\label{eq:ST}
\end{multline}
where both the electric and magnetic field correlation functions can
be written as derivatives of a vector potential operator
$\textbf{A}^E(\vec{r},\omega)$:
\begin{eqnarray}
\label{eq:EA}
E_i(\textbf{r},\omega) &=& -i\omega A^E_i(\textbf{r},\omega) \\
\label{eq:BA}
\mu H_i(\textbf{r},\omega) &=& (\nabla\times)_{ij} A^E_j(\textbf{r},\omega)
\end{eqnarray}
We explicitly place a superscript on the vector potential in order to
refer to our choice of gauge [\eqreftwo{EA}{BA}], in which $\vec{E}$
is obtained as a time-derivative of $\vec{A}$. The
fluctuation-dissipation theorem relates the correlation function of
$\textbf{A}^E$ to the photon Green's function
$G^E_{ij}(\omega;\textbf{r},\textbf{r}^\prime)$:
\begin{equation}
\langle A^E_i(\textbf{r},\omega)A^E_j(\textbf{r}^\prime,\omega)\rangle =
-\frac{\hbar}{\pi} \Im G^E_{ij}(\omega,\textbf{r},\textbf{r}^\prime),
\label{eq:FDT}
\end{equation}
where $G^E_{ij}$ is the vector potential $A^E_i$ in response to an
electric dipole current $\vec{J}$ along the $\hat{\vec{e}}_j$
direction:
\begin{equation}
\left[\nabla\times \frac{1}{\mu(\textbf{r},\omega)} \nabla\times{} -
\omega^2 \varepsilon(\textbf{r}, \omega) \right]
\textbf{G}^E_j(\omega;\textbf{r},\textbf{r}^\prime) =
\delta(\textbf{r}-\textbf{r}^\prime)\hat{\textbf{e}}_j,
\label{eq:GEOM}
\end{equation}
Given $G^E_{ij}$, one can use \eqreftwo{EA}{BA} in conjunction with
\eqref{FDT} to express the field correlation functions at points
$\vec{r}$ and $\vec{r}^\prime$ in terms of the photon Green's
function:
\begin{align}
\label{eq:EEG}
\langle E_i(\textbf{r},\omega) E_j(\textbf{r}^\prime,\omega)\rangle 
&= \frac{\hbar}{\pi}\omega^2 \Im G^E_{ij}(\omega;\textbf{r},\textbf{r}^\prime) \\
\label{eq:HHG}
\langle H_i(\textbf{r},\omega)H_j(\textbf{r}^\prime,\omega)\rangle &=
-\frac{\hbar}{\pi}(\nabla\times)_{il}(\nabla^\prime\times)_{jm} \Im G^E_{lm}(\textbf{r},\textbf{r}^\prime,\omega),
\end{align}

In order to find the force via \eqref{Force}, we must first compute
$G^E_{ij}(\vec{r},\vec{r}^\prime = \vec{r}, \omega)$ at every $\vec{r}$
on the surface of integration $S$, and for every
$\omega$~\cite{Lifshitz80}.  \Eqref{GEOM} can be solved numerically in a
number of ways, such as by a finite-difference
discretization~\cite{Rodriguez07:PRA}: this involves discretizing
space and solving the resulting matrix eigenvalue equation using
standard numerical linear algebra techniques~\cite{barrett94,
  Trefethen97}.  We note that finite spatial discretization
automatically regularizes the singularity in $G^E_{ij}$ at
$\vec{r}=\vec{r}'$, making $G^E_{ij}$ finite
everywhere~\cite{Rodriguez07:PRA}.

\subsection{Complex Frequency Domain}

The present form of \eqref{GEOM} is of limited computational utility
because it gives rise to an oscillatory integrand with non-negligible
contributions at all frequencies, making numerical integration
difficult~\cite{Rodriguez07:PRA}. However, the integral over $\omega$ can
be re-expressed as the imaginary part of a contour integral of an
analytic function by commuting the $\omega$ integration with the $\Im$
operator in \eqreftwo{EEG}{HHG}.  Physical causality implies that
there can be no poles in the integrand in the upper complex plane.
The integral, considered as a complex contour integral, is then
invariant if the contour of integration is deformed above the real
frequency axis and into the first quadrant of the complex frequency
plane, via some mapping $\omega \rightarrow \omega(\xi)$. This allows us
to add a positive imaginary component to the frequency, which causes
the force integrand to decay rapidly with increasing
$\xi$~\cite{RodriguezMc09:PRL}. In particular, upon deformation,
\eqref{GEOM} is mapped to:
\begin{equation}
\left[\nabla\times \frac{1}{\mu(\textbf{r},\omega)} \nabla\times{} -
\omega^2(\xi) \varepsilon(\textbf{r}, \omega) \right]
\textbf{G}^E_j(\xi;\textbf{r},\textbf{r}^\prime) =
\delta(\textbf{r}-\textbf{r}^\prime)\hat{\textbf{e}}_j,
\label{eq:cGEOM}
\end{equation}
and \eqreftwo{EEG}{HHG} are mapped to:
\begin{align}
\label{eq:EEG2}
\langle E_i(\textbf{r},\omega) E_j(\textbf{r}^\prime,\omega)\rangle 
&= \frac{\hbar}{\pi}\omega^2 G^E_{ij}(\omega;\textbf{r},\textbf{r}^\prime) \\
\label{eq:HHG2}
\langle H_i(\textbf{r},\omega)H_j(\textbf{r}^\prime,\omega)\rangle &=
-\frac{\hbar}{\pi}(\nabla\times)_{il}(\nabla^\prime\times)_{jm} G^E_{lm}(\textbf{r},\textbf{r}^\prime,\omega),
\end{align}
\Eqref{Force} becomes:
\begin{equation}
    F_i = \Im \int_0^\infty d\xi \frac{d\omega}{d\xi}
    \oiint_{\mathrm{surface}} \sum_j \langle T_{ij}(\vec{r},\omega)
    \rangle \, dS_j \, ,
\label{eq:newForce}
\end{equation}
[Note that a finite spatial grid (as used in the present approach)
requires no further regularization of the integrand, and the finite
value of all quantities means there is no difficulty in commuting the
$\Im$ operator with the integration.] 


We can choose from a general class of contours, provided that they
satisfy $\omega(0) = 0$ and remain above the real $\xi$ axis. The
standard contour $\omega(\xi) = i\xi$ is a Wick rotation, which is
known to yield a force integrand that is smooth and exponentially
decaying in $\xi$~\cite{Lifshitz80}. In general, the most suitable
contour will depend on the numerical method being employed.  A Wick
rotation guarantees a strictly positive-definite and real-symmetric
Green's function, making \eqref{GEOM} solvable by the most efficient
numerical techniques (e.g. the conjugate-gradient
method)~\cite{Trefethen97}.  One can also solve \eqref{GEOM} for
arbitrary $\omega(\xi)$~\cite{RodriguezMc09:PRL}, but this will
generally involve the use of direct solvers or more complicated
iterative techniques~\cite{barrett94}.  However, the class of contours
amenable to an efficient time-domain solution is more restricted. For
instance, a Wick rotation turns out to be unstable in the time domain
because it implies the presence of gain~\cite{RodriguezMc09:PRL}.

\subsection{Time Domain Approach}
\label{sec:time-domain}

It is possible to solve \eqref{GEOM} in the time domain by evolving
Maxwell's equations in response to a delta-function current impulse
$\vec{J}(\vec{r},t) = \delta(\vec{r}-\vec{r}') \delta(t -
t')\hat{\vec{e}}_j$ in the direction of $\hat{\vec{e}}_j$.  $G^E_{ij}$
can then be directly computed from the Fourier transform of the
resulting $\vec{E}$ field. However, obtaining a smooth and decaying
force integrand requires expressing the mapping $\omega \to
\omega(\xi)$ in the time-domain equations of motion.  A simple way to
see the effect of this mapping is to notice that~\eqref{cGEOM} can be
viewed as the Green's function at real ``frequency'' $\xi$ and complex
dielectric~\cite{RodriguezMc09:PRL}:
\begin{equation}
\varepsilon_c(\vec{r},\xi) = \frac{\omega^2(\xi)}{\xi^2} \varepsilon(\vec{r}) 
\label{eq:dispersion}
\end{equation}
where for simplicity we have taken $\mu$ and $\varepsilon$ to be
frequency-independent. We assume this to be the case for the remainder
of the manuscript.  At this point, it is important to emphasize that
the original physical system $\varepsilon$ at a frequency $\omega$ is
the one in which Casimir forces and fluctuations appear; the
dissipative system $\varepsilon_c$ at a frequency $\xi$ is merely an
artificial technique introduced to compute the Green's function.

\indent Integrating along a frequency contour $\omega(\xi)$ is
therefore equivalent to making the medium dispersive in the form of
\eqref{dispersion}.  Consequently, the time domain equations of motion
under this mapping correspond to evolution of the fields in an
effective dispersive medium given by $\varepsilon_c(\vec{r},\xi)$.

To be suitable for FDTD, this medium should have three properties: it
must respect causality, it cannot support gain (which leads to
exponential blowup in time-domain), and it should be easy to
implement.  A Wick rotation is very easy to implement in the
time-domain: it corresponds to setting $\varepsilon_c = -\varepsilon$.
However, a negative epsilon represents gain (the refractive index is
$\pm \sqrt{\varepsilon}$, where one of the signs corresponds to an
exponentially growing solution).  We are therefore forced to consider
a more general, frequency-dependent $\varepsilon_c$.

Implementing arbitrary dispersion in FDTD generally requires the
introduction of auxiliary fields or higher order time-derivative terms
into Maxwell's equations, and can in general become computationally
expensive~\cite{Taflove00}. The precise implementation will depend
strongly on the choice of contour $\omega(\xi)$. However, almost any
dispersion will suit our needs, as long as it is causal and
dissipative (excluding gain).  A simple choice is an
$\varepsilon_c(\vec{r},\xi)$ corresponding to a medium with
frequency-independent conductivity $\sigma$:
\begin{equation}
\varepsilon_c(\vec{r},\xi) = \varepsilon(\vec{r}) \left( 1 + \frac{i \sigma}{\xi}\right)
\end{equation}

This has three main advantages: first, it is implemented in many FDTD
solvers currently in use; second, it is numerically stable; and third,
it can be efficiently implemented without an auxiliary differential
equation~\cite{Taflove00}.  In this case, the equations of motion in
the time domain are given by:
\begin{eqnarray}
\label{eq:FDTD1}
\frac{\partial \mu\textbf{H}}{\partial t} &=& -\nabla \times \vec{E}
\\ \frac{\partial \varepsilon \vec{E}}{\partial t} &=& \nabla \times \vec{H} -
\sigma \varepsilon \vec{E} - \vec{J}
\label{eq:FDTD2}
\end{eqnarray}
Writing the conductivity term as $\sigma \varepsilon$ is slightly
nonstandard, but is convenient here for numerical reasons.  In
conjunction with~\eqreftwo{EA}{BA}, and a Fourier transform in $\xi$,
this yields a photon Green's function given by:
\begin{equation}
\left[\nabla\times \frac{1}{\mu(\textbf{r})} \nabla\times{} -
\xi^2 \varepsilon(\textbf{r}) \left(1+\frac{i\sigma}{\xi}\right)
\right]
\textbf{G}_j(\xi;\textbf{r},\textbf{r}^\prime) =
\delta(\textbf{r}-\textbf{r}^\prime)\hat{\textbf{e}}_j,
\label{eq:sigma-EOM}
\end{equation}
\indent This corresponds to picking a frequency contour of the form:
\begin{equation}
\omega(\xi) \equiv \xi \sqrt{1+\frac{i\sigma}{\xi}},
\label{eq:Contour}
\end{equation}
Note that, in the time domain, the frequency of the fields is $\xi$,
and not $\omega$, i.e. their time dependence is $e^{-i\xi t}$. The
only role of the conductivity $\sigma$ here is to introduce an
imaginary component to \eqref{sigma-EOM} in correspondence with a
complex-frequency mapping.  It also explicitly appears in the final
expression for the force, \eqref{newForce}, as a multiplicative
(Jacobian) factor.

The standard FDTD method involves a discretized form of
\eqreftwo{FDTD1}{FDTD2}, from which one obtains $\textbf{E}$ and
$\textbf{B}$, not $G^E_{ij}$.  However, in the frequency domain, the
photon Green's function, being the solution to~\eqref{GEOM}, solves
exactly the same equations as those satisfied by the electric field
$\textbf{E}$, except for a simple multiplicative factor
in~\eqref{EA}. Specifically, $G^E_{ij}$ is given in terms of $\vec{E}$
by:
\begin{equation}
G^E_{ij}(\xi;\textbf{r},\textbf{r}^\prime) =
-\frac{E_{i,j}(\textbf{r},\xi)}{i\xi \mathcal{J}(\xi)},
\label{eq:Gij}
\end{equation}
where $E_{i,j}(\vec{r}, \xi)$ denotes the field in the $i$th direction
due to a dipole current source
$\textbf{J}(\textbf{r},t)=\mathcal{J}(t)\delta(\textbf{r}-\textbf{r}^\prime)
\hat{\mathrm{e}}_j$ placed at $\textbf{r}^\prime$ with time-dependence
$\mathcal{J}(t)$, e.g. $\mathcal{J}(t) = \delta(t)$. 

In principle, we can now compute the electric- and magnetic-field
correlation functions by using~\eqreftwo{EEG2}{HHG2}, with
$\omega(\xi)$ given by \eqref{Contour}, and by setting $\vec{r} =
\vec{r}'$ in \eqref{HHG2}.  Since we assume a discrete spatial grid,
no singularities arise for $\vec{r}=\vec{r}'$, and in fact any
$\vec{r}$-independent contribution is canceled upon integration over
$S$. This is straightforward for \eqref{EEG}, since the
$\vec{E}$-field correlation function only involves a simple
multiplication by $\omega^2(\xi)$. However, the $\vec{H}$-field
correlation function, \eqref{HHG}, involves derivatives in space.
Although it is possible to compute these derivatives numerically as
finite differences, it is conceptually much simpler to pick a
different vector potential, analogous to~\eqreftwo{EA}{BA}, in which
$\vec{H}$ is the time-derivative of a vector potential $\vec{A}^H$.
As discussed in the Appendix, this choice of vector potential implies
a frequency-independent magnetic conductivity, and a magnetic, instead
of electric, current.  The resulting time-domain equations of motion
are:
\begin{eqnarray}
\label{eq:FDTD1d}
\frac{\partial \mu\vec{H}}{\partial t} &=& -\nabla \times \vec{E} + \sigma \mu \vec{H} - \vec{J} \\
\frac{\partial \varepsilon\vec{E}}{\partial t} &=& \nabla \times \vec{H}
\label{eq:FDTD2d}
\end{eqnarray}
In this gauge, the new photon Green's function $G_{ij}^H = \langle
A_i^H(\vec{r},\xi) A_j^H(\vec{r}',\xi)\rangle$ and the field $\vec{H}$
in response to the current source $\vec{J}$ are related by:
\begin{equation}
G^H_{ij}(\xi; \textbf{r}, \textbf{r}^\prime) =
-\frac{H_{i,j}(\textbf{r},\xi)}{i\xi \mathcal{J}(\xi)},
\label{eq:Gdual}
\end{equation}
where the magnetic-field correlation function:
\begin{equation}
\langle H_i(\textbf{r},\xi)H_j(\textbf{r}^\prime,\xi)\rangle 
= \frac{\hbar}{\pi} \omega^2(\xi) G^H_{ij}(\xi; \textbf{r},\textbf{r}^\prime),
\label{eq:HHGH}
\end{equation}
is now defined as a frequency multiple of $G^H_{ij}$ rather than by a
spatial derivative of $G^E_{ij}$.

This approach to computing the magnetic correlation function has the
advantage of treating the electric and magnetic fields on the same
footing, and also allows us to examine only the field response at the
location of the current source. The removal of spatial derivatives
also greatly simplifies the incorporation of discretization into our
equations (see Appendix for further discussion).  The use of magnetic
currents and conductivities, while unphysical, are easily implemented
numerically.  Alternatively, one could simply interchange
$\varepsilon$ and $\mu$, $\vec{E}$ and $\vec{H}$, and run the
simulation entirely as in~\eqreftwo{FDTD1}{FDTD2}.

The full force integral is then expressed in the symmetric form:
\begin{equation}
F_i = \Im \frac{\hbar}{\pi} \int_{-\infty}^\infty d\xi ~g(\xi)  
\left( \Gamma^E_i(\xi) + \Gamma^H_{i}(\xi)\right),
\label{eq:Force_fields}
\end{equation}
where
\begin{eqnarray}
\label{eq:Gamma1}
\Gamma^E_i(\xi) &\equiv& \oiint_S \sum_j \varepsilon(\vec{r}) \left( E_{i,j}(\vec{r}) - \frac{1}{2}\delta_{ij}\sum_k E_{k,k}(\vec{r}) \right) \, dS_j \,\\
\Gamma^H_{i}(\xi) &\equiv& \oiint_S \sum_j \frac{1}{\mu(\vec{r})}\left( H_{i,j}(\vec{r}) - \frac{1}{2}\delta_{ij}\sum_k H_{k,k}(\vec{r}) \right) \, dS_j \,
\label{eq:Gamma2}
\end{eqnarray}
represent the surface-integrated field responses in the frequency
domain, with $E_{i,j}(\vec{r}) \equiv E_{i,j}(\vec{r}; \xi)$. For
notational simplicity, we have also defined:
\begin{equation}
g(\xi) \equiv \frac{\omega^2}{i \xi \mathcal{J}(\xi)} \frac{d\omega}{d\xi} \Theta(\xi)
\label{eq:gw}
\end{equation}
Here, the path of integration has been extended to the entire real
$\xi$-axis with the use of the unit-step function $\Theta(\xi)$ for
later convenience.

The product of the fields with $g(\xi)$ naturally decomposes the
problem into two parts: computation of the surface integral of the
field correlations $\Gamma$, and of the function $g(\xi)$.  The
$\Gamma_i$ contain all the structural information, and are
straightforward to compute as the output of any available FDTD solver
with no modification to the code. This output is then combined with
$g(\xi)$, which is easily computed analytically, and integrated
in~\eqref{Force_fields} to obtain the Casimir force.  As discussed in
\secref{discretization}, the effect of spatial and temporal
discretization enters explicitly only as a slight modification to
$g(\xi)$ in~\eqref{Force_fields}, leaving the basic conclusions
unchanged.

\subsection{Evaluation in the Time Domain}

It is straightforward to evaluate~\eqref{Force_fields} in the
frequency domain via a dipole current $\mathcal{J}(t) = \delta(t)$,
which yields a constant-amplitude current $\mathcal{J}(\xi) = 1$.
Using the frequency-independent conductivity contour \eqref{Contour},
corresponding to \eqreftwo{FDTD1}{FDTD2}, we find the following
explicit form for $g(\xi)$:
\begin{equation}
g(\xi) = -i\xi \left(1 + \frac{i \sigma}{\xi}\right) \frac{1+i\sigma / 2 \xi}{\sqrt{1+i\sigma /
    \xi}} \Theta(\xi)
\label{eq:gw-delta}
\end{equation}
One important feature of \eqref{gw-delta} is that $g(\xi) \to
\sqrt{i\sigma^3/\xi}$ becomes singular in the limit as $\xi\to 0$.
Assuming that $\Gamma^E(\xi)$ and $\Gamma^H(\xi)$ are continuous at
$\xi=0$ (in general they will not be zero), this singularity is
integrable.  However, it is cumbersome to integrate in the frequency
domain, as it requires careful consideration of the time window for
calculation of the field Fourier transforms to ensure accurate
integration over the singularity.

As a simple alternative, we use the convolution theorem to re-express
the frequency ($\xi$) integral of the product of $g(\xi)$ and
$\Gamma^E(\xi)$ arising in \eqref{Force_fields} as an integral over
time $t$ of their Fourier transforms $g(-t)$ and
$\Gamma^E(t)$. Technically, the Fourier transform of $g(\xi)$ does not
exist because $g(\xi) \sim \xi$ for large $\xi$. However, the integral
is regularized below using the time discretization, just as the
Green's function above was regularized by the spatial discretization.
(As a convenient abuse of notation, $\xi$ arguments will always denote
functions in the frequency domain, and $t$ arguments their Fourier
transforms in the time domain.)

Taking advantage of the causality conditions
($\Gamma^E(t),\,\Gamma^H(t) = 0$ for $t < 0$) yields the following
expression for the force expressed purely in the time domain:
\begin{equation}
F_i = \Im \frac{\hbar}{\pi}\int_0^\infty dt ~g(-t) \left(\Gamma^E_i(t) + \Gamma^H_i(t)\right)
\label{eq:time-force}
\end{equation}

The advantage of evaluating the force integral in the time domain is
that, due to the finite conductivity and lack of sources for $t > 0$,
$\Gamma(t)$ will rapidly decay in time.  As will be shown in the next
section, $g(-t)$ also decays with time.  Hence, although dissipation
was originally introduced to permit a natural high-frequency cutoff to
our computations, it also allows for a natural time cutoff $T$.  We
pick $T$ such that, for times $t > T$, knowledge of the fields will
not change the force result in~\eqref{time-force} beyond a
predetermined error threshold.  This approach is very general as it
requires no precise knowledge of how the fields decay with time.

\subsection{Properties of \emph{g}($\mathbf{-}$\emph{t})}

Given $g(\xi)$, the desired function $g(-t)$ is a Fourier
transform. However, the discretization of time in FDTD implies that
the frequency domain becomes periodic and that $g(t) = g(n\Delta t)$
are actually Fourier series coefficients, given by:
\begin{equation}
  g(n\Delta t) = \int_0^{2\pi/\Delta t} d\xi \, g_d(\xi) e^{-i \xi n \Delta t},
\label{eq:gndt}
\end{equation}
where $g_d(\xi)$ is the discretized form of \eqref{gw} and is given in
the Appendix by \eqref{gwd}. These Fourier series coefficients are
computed by a sequence of numeric integrals that can be evaluated in a
variety of ways. It is important to evaluate them accurately in order
to resolve the effect of the $\xi=0$ singularity. For example, one
could use a Clenshaw-Curtis scheme developed specifically for Fourier
integrals~\cite{piessens83}, or simply a trapezoidal rule with a large
number of points that can be evaluated relatively quickly by an FFT
(e.g. for this particular $g(\xi)$, $10^7$ points is sufficient).

\begin{figure}[tb]
\includegraphics[width=0.46\textwidth,height=0.33\textwidth]{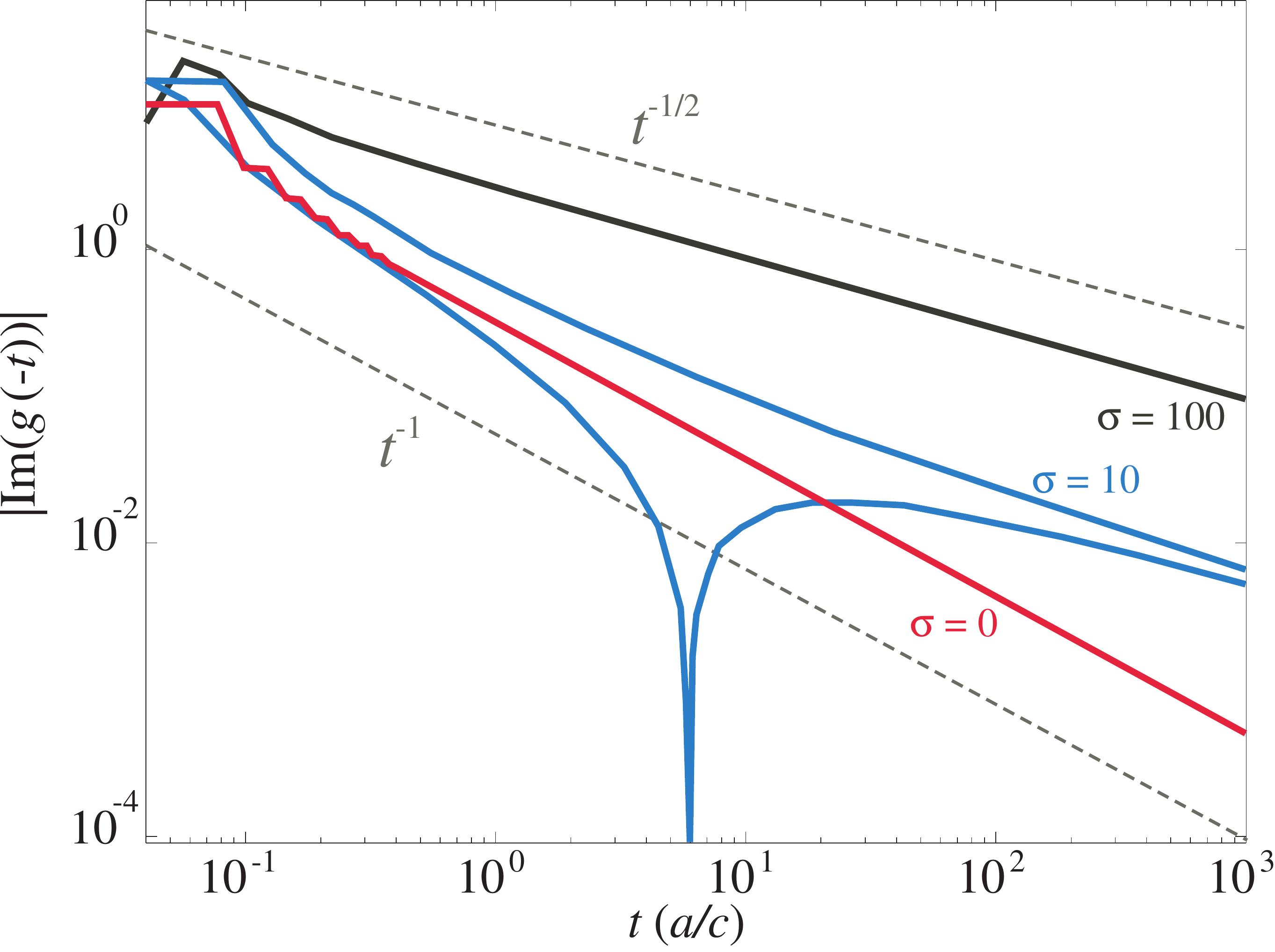}
\caption{$|\Im g(t)|$ for various values of $\sigma$, illustrating the
  transition from $t^{-1}$ to $t^{-1/2}$ power-law decay as $\sigma$
  increases.  Because there are strong oscillations in $g(t)$ at the
  Nyquist frequency for intermediate $\sigma$, for clarity we plot the
  positive and negative terms in $g(t)$ as separate components.}
\label{fig:gt}
\end{figure}

Since it is possible to employ strictly-real current sources in FDTD,
giving rise to real $\Gamma$, and since we are only interested in
analyzing the influence of $g(t)$ on \eqref{time-force}, it suffices
to look at $\Im g(-t)$. Furthermore, $g(t)$ will exhibit rapid
oscillations at the Nyquist frequency due to the delta-function
current, and therefore it is more convenient to look at its absolute
value. \Figref{gt}, below, plots the envelope of $|\Im g(-t)|$ as a
function of $t$, where again, $g(t)$ is the Fourier transform of
\eqref{gw}.

As anticipated in the previous section, $g(t)$ decays in time.
Interestingly, it exhibits a transition from $\sim t^{-1}$ decay at
$\sigma = 0$ to $\sim t^{-1/2}$ decay for large $\sigma$.  The slower
decay at long times for larger $\sigma$ arises from a transition in
the behavior of~\eqref{gw-delta} from the singularity at $\xi=0$.

\section{Properties of the Method}

In this section we discuss the practical implementation of the
time-domain algorithm (using a freely-available time domain
solver~\cite{Farjadpour06} that required no modification).  We analyze
its properties applied to the simplest parallel-plate geometry
[\figref{1d-plates}], which illustrate the essential features in the
simplest possible context.  In particular, we analyze important
computational properties such as the convergence rate and the impact
of different conductivity choices.  Part~II of this manuscript, in
preparation, demonstrates the method for more complicated two- and
three-dimensional geometries~\cite{McCauleyRo09}.

\subsection{Fields in Real Time}

The dissipation due to positive $\sigma$ implies that the fields, and
hence $\Gamma^E(t)$, will decay exponentially with time.  Below, we
use a simple one-dimensional example to understand the consequences of
this dissipation for both the one-dimensional parallel plates and the
two-dimensional piston configuration.  The simplicity of the
parallel-plate configuration allows us to examine much of the behavior
of the time-domain response analytically.  (The understanding gained
from the one-dimensional geometry can be applied to higher
dimensions.)  Furthermore, we confirm that the error in the Casimir
force due to truncating the simulation at finite time decreases
exponentially (rather than as $t^{-1}$, as it would for no
dissipation).

\subsubsection{One-dimensional Parallel Plates}

\begin{figure}[tb]
\includegraphics[width=0.48\textwidth]{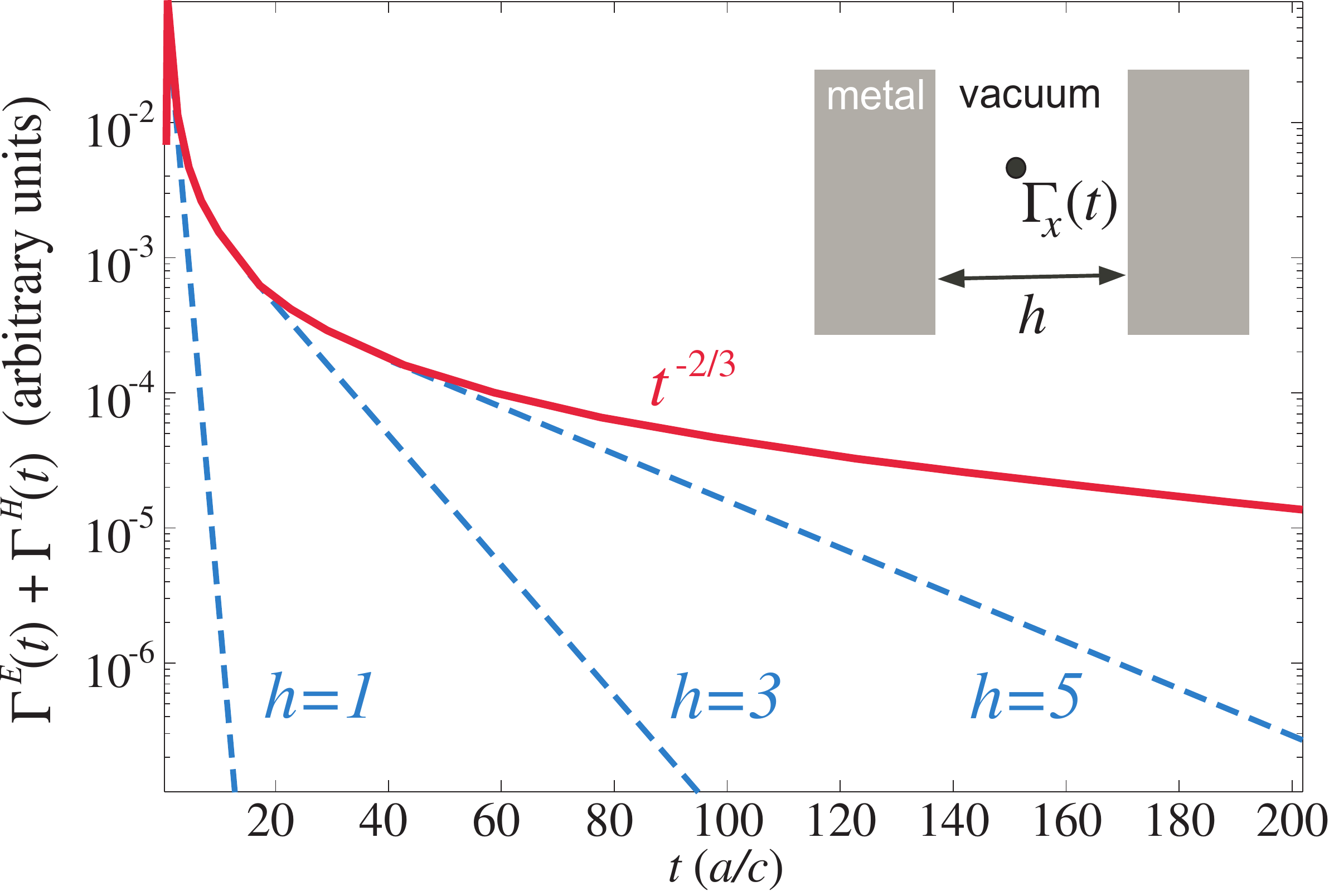}
\caption{$\Gamma^E_{x}(t)+\Gamma^H_{x}(t)$ for a set of
  one-dimensional parallel plates as the separation $h$ is varied.
  The inset shows the physical setup.}
\label{fig:1d-plates}
\end{figure}

To gain a general understanding of the behavior of the system in the
time domain, we first examine a simple configuration of perfectly
metallic parallel plates in one dimension.  The plates are separated
by a distance $h$ (in units of an arbitrary distance $a$) in the $x$
dimension, as shown by the inset of~\figref{1d-plates}.  The figure
plots the field response $\Gamma^E_{x}(t) + \Gamma^H_x(t)$, in
arbitrary units, to a current source $\mathcal{J}(t)=\delta(t)$ for
increasing values of $h$, with the conductivity set at $\sigma = 10 \,
(2\pi c / a)$ .

\Figref{1d-plates} shows the general trend of the field response as a
function of separation.  For short times, all fields follow the same
power-law envelope, and later rapidly transition to exponential decay.
Also plotted for reference is a $t^{-3/2}$ curve, demonstrating that
the envelope is in fact a power law.

We can understand the power law envelope by considering the vacuum
Green's function $G^E$ in the case $h\to\infty$ (analogous conclusions
hold for $G^H$).  In the case $h\to\infty$, one can easily solve for
the vacuum Green's function $G^E(\xi,\vec{r}-\vec{r}')$ in one
dimension for real frequency $\xi$:
\begin{equation}
G^E(\xi,\vec{r}-\vec{r}') = \frac{e^{i\xi|\vec{r}-\vec{r}'|}}{i\xi}
\end{equation}

We then analytically continue this expression to the complex frequency
domain via~\eqref{Contour} and compute the Fourier transform $\int
d\xi e^{i\xi t}G^E(\omega(\xi))$.  Setting $\vec{r}=\vec{r}'$ in the
final expression, one finds that, to leading order, $G^E(t)\sim
t^{-3/2}$.  This explains the behavior of the envelope
in~\figref{1d-plates} and the short-time behavior of the Green's
functions: it is the field response of vacuum.

Intuitively, the envelope decays only as a power in $t$ because it
receives contributions from a continuum of modes, all of which are
individually decaying exponentially (this is similar to the case of
the decay of correlations in a thermodynamic system near a critical
point~\cite{Goldenfeld92}).  For a finite cavity, the mode spectrum is
discrete --- the poles in the Green's function of the non-dissipative
physical system are pushed below the real frequency axis in this
dissipative, unphysical system, but they remain discretely spaced.

At short times, the field response of a finite cavity will mirror that
of an infinite cavity because the fields have not yet propagated to
the cavity walls and back.  As $t$ increases, the cavity response will
transition to a discrete sum of exponentially decaying modes.
From~\eqref{Contour}, higher-frequency modes have a greater
imaginary-frequency component, so at sufficiently long times the
response will decay exponentially, the decay being determined by the
lowest-frequency cavity mode.  The higher the frequency of that mode,
the faster the dissipation.

This prediction is confirmed in~\figref{1d-plates}: as $h$ decreases,
the source ``sees'' the walls sooner.  From the standpoint of
computational efficiency, this method then works best when objects are
in close proximity to one another (although not so close that spatial
resolution becomes an issue), a situation of experimental interest.

\subsubsection{Convergence of the Force}

We now examine the force on the parallel plates.  From the above
discussions of the field decay and the decay of $g(t)$, we expect the
time integral in~\eqref{time-force} to eventually converge
exponentially as a function of time.  In the interest of quantifying
this convergence, we define the time dependent ``partial force''
$F_i(t)$ as:
\begin{equation}
F_i(t) \equiv \Im \frac{\hbar}{\pi}\int_0^t dt^\prime ~g(-t^\prime) \left(\Gamma^E_i(t^\prime)+\Gamma^H_i(t^\prime)\right)
\label{eq:partial-force}
\end{equation}

Letting $F_i(\infty)$ denote the $t\to\infty$ limit of $F_i(t)$, which
is the actual Casimir force, we define the relative error
$\Delta_i(t)$ in the $i$-th component of the force as:
\begin{equation}
\Delta_i(t) \equiv \left| \frac{ F_i(t) - F_i(\infty)}{F_i(\infty)}\right|
\end{equation}

\begin{figure}[tb]
\includegraphics[width=0.45\textwidth]{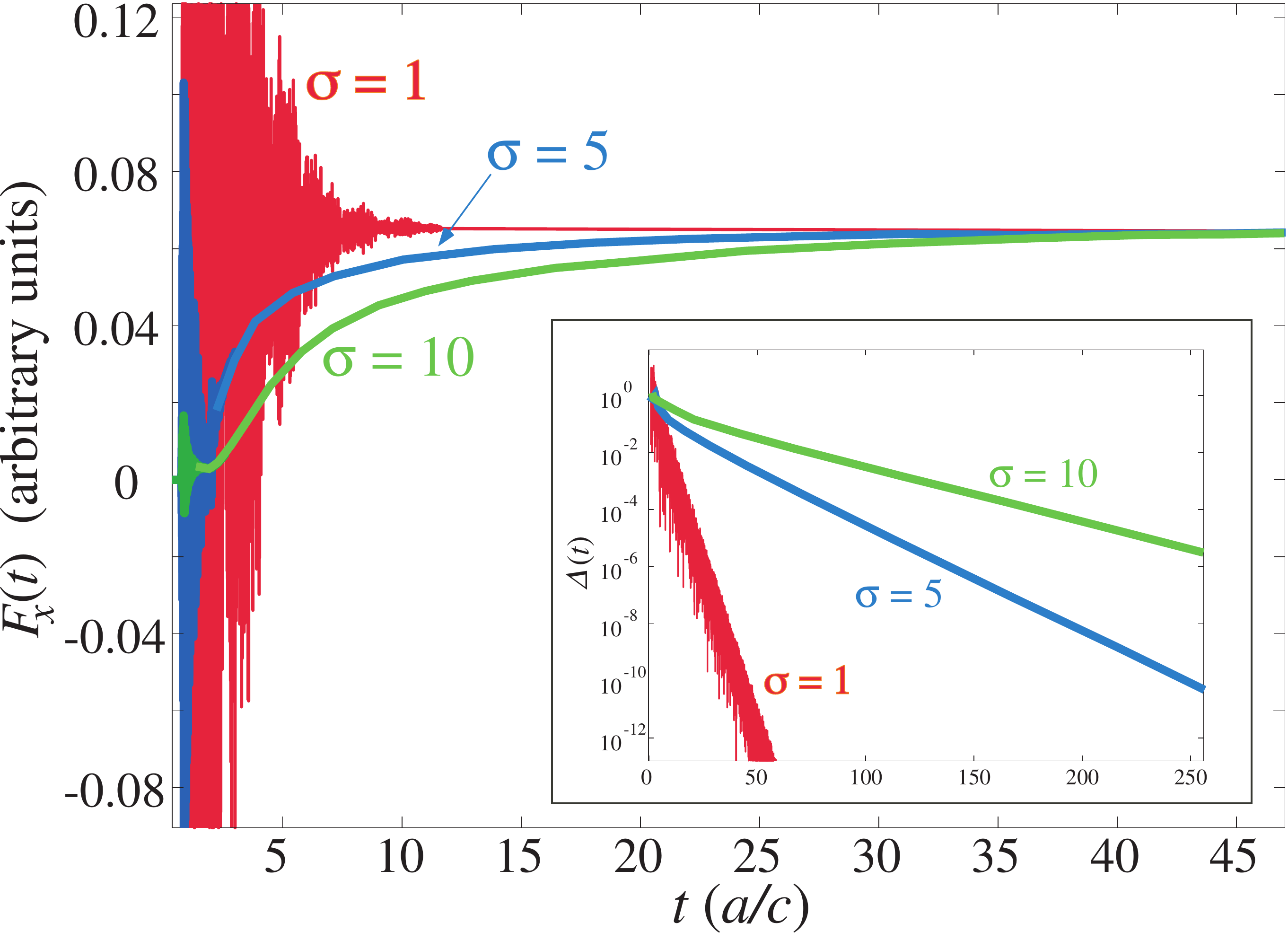}
\caption{Partial force as defined in~\eqref{partial-force} for
  one-dimensional parallel plates as a function of time $t$. (Inset):
  Relative error $\Delta (t)$ as a function of $t$ on a semi log
  scale.}
\label{fig:force-1d}
\end{figure}

We plot $F_x(t)$ in \figref{force-1d} for the one-dimensional
parallel-plate structure with different values of $\sigma$.  The inset
plots $\Delta(t)$ for the same configuration.  As expected, the
asymptotic value of $F_x(t)$ is independent of $\sigma$, and
$\Delta(t)$ converges exponentially to zero.

For $\sigma$ near zero, the force is highly oscillatory.  In one
dimension this gives the most rapid convergence with time, but it is
problematic in higher dimensions.  This is because, in
higher-dimensional systems, $S$ consists of many points, each
contributing a response term as in~\figref{force-1d}.  If $\sigma$ is
small, every one of these terms will be highly oscillatory, and the
correct force~\eqref{partial-force} will only be obtained through
delicate cancellations at all points on $S$.  Small $\sigma$ is thus
very sensitive to numerical error. 

Increasing $\sigma$ smooths out the response functions, as higher
frequency modes are damped out.  However, somewhat counterintuitively,
it also has the effect of slowing down the exponential convergence.
One can understand the asymptotic behavior of the force by considering
the equations of motion~\eqref{sigma-EOM} as a function of $\sigma$
and $\xi$.  When the response function exhibits few if any
oscillations we are in the regime where $\sigma \gg \xi$.  In this
limit, the approximate equations of motion are:

\begin{equation}
\left[\nabla\times \frac{1}{\mu(\textbf{r})} \nabla\times{} -
  i\sigma \xi \varepsilon(\textbf{r}) \right]
\textbf{G}_j(\xi;\textbf{r},\textbf{r}^\prime) =
\delta(\textbf{r}-\textbf{r}^\prime)\hat{\textbf{e}}_j
\label{eq:large-sigma}
\end{equation}

In the limit of \eqref{large-sigma}, the eigenfrequency $\xi$ of a
given spatial mode scales proportional to $-i/\sigma$.  The
lowest-frequency mode therefore has a time-dependence $\sim e^{-C
  t/\sigma}$, for some constant $C > 0$.  Since the decay of the force
at long times is determined by this mode, we expect the decay time to
scale inversely with $\sigma$ in the limit of very high $\sigma$.
This is suggested in~\figref{force-1d} and confirmed by further
numerical experiments.

Additionally, from~\eqref{large-sigma} we see that in the case of a
homogeneous one-dimensional cavity, the solutions have a quadratic
dispersion $\xi \sim ik^2$, for spatial dependence $e^{ikx}$, and so
the lowest cavity frequency scales as the inverse square of the cavity
size.  This means that the rate of exponential convergence
of~\figref{1d-plates} should vary as $\sim h^{-2}$ in the limit of
very large $\sigma$.  This scaling is approximately apparent from
\figref{1d-plates}, and further experiments for much larger $\sigma$
confirm the scaling.  We thus see that in this limit, the effect of
increasing $\sigma$ by some factor is analogous to increasing the wall
spacing of the cavity by the square root of that factor.

The present analysis shows that there are two undesirable extremes.
When $\sigma$ is small, rapid oscillations in $F_i(t)$ will lead to
large numerical errors in more than one dimension.  When $\sigma$ is
large, the resulting frequency shift will cause the cavity mode to
decay more slowly, resulting in a longer run time.  The optimal
$\sigma$ lies somewhere in between these two extremes and will
generally depend on the system being studied.  For the systems
considered in this paper, with a typical scale $\approx a$, $\sigma
\sim 1 \, (2\pi c / a)$ appears to be a good value for efficient and
stable time-domain computation.

\section{Concluding Remarks}

An algorithm to compute Casimir forces in FDTD has several practical
advantages.  FDTD algorithms that solve Maxwell's equations with
frequency-independent conductivity, and even more complicated
dispersions, are plentiful and well-studied.  They are stable,
convergent, and easily parallelized.  Although the current formulation
of our method requires the evaluation of $G_{ij}(\vec{r})$ along a
surface $S$, requiring a separate calculation of the fields for each
dipole source in $S$, all of these sources can be simulated in
parallel, with no communication between different simulations until
the very end of the computation.  In addition, many FDTD solvers will
allow the computational cell for each source to be parallelized,
providing a powerful method capable of performing large computations.

The calculations of this paper employed non-dispersive materials in
the original ($\omega$) system.  However, the theoretical analysis
applies equally well to materials of arbitrary dispersion.  Any
materials that can be implemented in an FDTD solver (e.g. a sum of
Lorentzian dielectric resonances~\cite{Taflove00}) can also be
included, and existing algorithms have demonstrated the ability to
model real materials~\cite{Taflove00,YoungNe01}. Existing FDTD
implementations also handle anisotropy in $\varepsilon$ and $\mu$,
multiple types of boundary conditions, and other
complications~\cite{Taflove00}.

In principle, the computational scaling of this FDTD method is
comparable to finite-difference frequency-domain (FDFD)
methods~\cite{Rodriguez07:PRA}. In both cases, each solver step
(either a time step for FDTD or an iterative-solver step for FDFD)
requires $O(N)$ work for $N$ grid points. The number of time steps
required by an FDTD method is proportional to the diameter of the
computational cell, or $N^{1/d}$ in $d$ dimensions. With an ideal
multigrid solver, FDFD can in principle be solved by $O(1)$ solver
steps, but a simpler solver like conjugate gradient requires a number
of steps proportional to the diameter as
well~\cite{Rodriguez07:PRA}. In both cases, the number of points to be
solver on the surface $S$ is $O(N^{1-1/d})$. Hence, the overall
complexity of the simplest implementations (not multigrid) is
$O(N^2)$. We believe that future boundary-element
methods~\cite{Rodriguez07:PRA,ReidRo09} will achieve better
efficiency, but such methods require considerable effort to implement
and their implementation is specific to the homogeneous-medium Green's
function, which depends on the boundary conditions, dimensionality and
types of materials considered~\cite{Rao99}.

Part~II of this manuscript~\cite{McCauleyRo09}, in preparation, will
illustrate the method in various non-trivial two- and
three-dimensional geometries, including dispersive dielectrics.  In
addition, we introduce an optimization of our method (based on a
rapidly converging series expansion of the fields) that greatly speeds
up the spatial integral of the stress tensor. We also compute forces
in three-dimensional geometries with cylindrical symmetry, which
allows us to take advantage of the cylindrical coordinates support in
existing FDTD software~\cite{Farjadpour06} and employ a
two-dimensional computational cell.

\section*{ACKNOWLEDGEMENTS}

We would like to thank Peter Bermel and Ardavan Farjadpour for useful
discussions.  This work was supported by the Army Research Office
through the ISN under Contract No. W911NF-07-D-0004, the MIT Ferry
Fund, and by US DOE Grant No. DE-FG02-97ER25308 (ARW).

\section*{APPENDIX}

\subsection{Effects of Discretization}
\label{sec:discretization}

FDTD algorithms approximate both time and space by a discrete uniform
mesh.  Bearing aside the standard analysis of stability and
convergence~\cite{Taflove00}, this discretization will slightly modify
the analysis in the preceding sections. In particular, the use of a
finite temporal grid (resolution $\Delta t$) implies that all
continuous time derivatives are now replaced by a finite-difference
relation, which is most commonly taken to be a center difference:
\begin{equation}
\frac{\partial f}{\partial t} \approx \frac{f_i(\textbf{r},t + \Delta
  t/2) - f_i(\textbf{r},t - \Delta t/2)}{\Delta t} \equiv \partial^{(d)}_t f
\end{equation}
where $f(t)$ is an arbitrary function of time. The effect of temporal
discretization is therefore to replace the linear operator
$\partial/\partial t$ with $\partial^{(d)}_t$. The representation of this
operator is simple to compute in the frequency domain. Letting
$\partial^{(d)}_t$ act on a Fourier component of of $f(t)$ yields:
\begin{equation}
\partial^{(d)}_t e^{-i\xi t} = -i\xi_d e^{-i \xi t},
\end{equation}
where
\begin{equation}
\xi_d(\xi) \equiv \frac{2}{\Delta t} \sin\left(\frac{\xi \Delta t}{2}\right) e^{-i\frac{\xi \Delta t}{2}}
\label{eq:wd}
\end{equation}

The effect of discretization on the system is thus to replace $i \xi$
by $i \xi_d$ in the derivatives, which correspond to numerical
dispersion arising from the ultraviolet (Nyquist) frequency cutoff
$\pi/\Delta t$. Note that $\xi$ is still the frequency parameter
governing the time dependence of the Fourier components of $f(t)$ and
$\xi_d \to \xi$ in the limit of infinite resolution ($\Delta t \to
0$).


Because FDTD is convergent [$\xi_d = \xi + O(\Delta t^2)$], most of
the analysis can be done (as in this paper) in the $\Delta t \to 0$
limit. However, care must be taken in computing $g(t)$ because the
Fourier transform of $g(\xi)$, \eqref{gw}, does not exist as $\Delta t
\to 0$. We must compute it in the finite $\Delta t$ regime.  In
particular, the finite resolution requires, via \eqref{wd}, that we
replace $g(\omega)$ in \eqref{gw} by:
\begin{equation}
g_d(\xi) \equiv \frac{\omega_d^2}{i \xi_d \mathcal{J}(\xi)} \frac{d\omega}{d\xi}
\label{eq:gwd}
\end{equation}
Note that the Jacobian factor $d\omega / d\xi$ involves $\omega$ and
$\xi$, not $\omega_d$ and $\xi_d$, although of course the latter
converges to the former for $\Delta t \to 0$.  The basic principle is
that one must be careful to use the discrete analogues to continuous
solutions in cases where there is a divergence or regularization
needed.  This is the case for $g(\xi)$, but not for the Jacobian.

Similarly, if one wished to subtract the vacuum Green's function from
the Green's function, one needs to subtract the vacuum Green's
function as computed in the discretized vacuum.  Such a subtraction is
unnecessary if the stress tensor is integrated over a closed surface
(vacuum contributions are constants that integrate to zero), but is
useful in cases like the parallel plates considered here.  By
subtracting the (discretized) vacuum Green's function, one can
evaluate the stress tensor only for a single point between the plates,
rather than for a ``closed surface'' with another point on the other
side of the plates~\cite{Lifshitz80}.

As was noted before \eqref{gndt}, the Nyquist frequency $\pi/\Delta t$
regularizes the frequency integrations, similar to other ultraviolet
regularization schemes employed in Casimir force
calculations~\cite{lifshitz1, Mazzitelli06}. Because the total
frequency integrand in \eqref{Force} goes to zero for large $\xi$ (due
to cancellations occurring in the spatial integration and also due to
the dissipation introduced in our approach), the precise nature of this
regularization is irrelevant as long as $\Delta t$ is sufficiently
small (i.e., at high enough resolution).


\subsection{The Magnetic Correlation Function}

One way to compute the magnetic correlation function is by taking
spatial derivatives of the electric Green's function by \eqref{HHG},
but this is inefficient because a numerical derivative involves
evaluating the electric Green's function at multiple points. Instead,
we compute the magnetic Green's function directly, finding the
magnetic field in response to a magnetic current.  This formulation,
however, necessitates a change in the choice of vector
potentials~\eqreftwo{EA}{BA} as well as a switch from an electric to
magnetic conductivity, for reasons explained in this section.

\Eqreftwo{EA}{BA} express the magnetic field $\vec{B}$ as the curl of
the vector potential $\vec{A}^E$, enforcing the constraint that
$\vec{B}$ is divergence-free (no magnetic charge).  However, this is
no longer true when there is a magnetic current, as can be seen by
taking the divergence of both sides of Faraday's law with a magnetic
current $\vec{J}$, $\partial\vec{B}/\partial t = -\nabla\times\vec{E}
- \vec{J}$, since $\nabla\cdot\vec{J} \neq 0$ for a point-dipole
current $\vec{J}$. Instead, since there need not be any free electric
charge in the absence of an electric current source, one can switch to
a new vector potential $\vec{A}^H$ such that
\begin{align}
\label{eq:EA2}
  \varepsilon E_i(\vec{r},\omega) &= \left(\nabla \times \right)_{ij} A^H_j(\vec{r},\omega) \\
  H_i(\vec{r},\omega) &= -i\omega A^H_i(\vec{r},\omega).
\label{eq:BA2}
\end{align}
The desired correlation function is then given, analogous to \eqref{EEG}, by
\begin{equation}
\langle H_i(\textbf{r},\omega) H_j(\textbf{r}^\prime,\omega)\rangle 
= \frac{\hbar}{\pi}\omega^2 \Im G^H_{ij}(\omega;\textbf{r},\textbf{r}^\prime),
\end{equation}
where the photon magnetic Green's function $G^H$ solves [similar to
  \eqref{GEOM}]
\begin{equation}
\left[\nabla\times \frac{1}{\varepsilon(\textbf{r},\omega)} \nabla\times{} -
  \omega^2 \mu(\textbf{r},\omega)\right]
\textbf{G}^H_j(\omega;\textbf{r},\textbf{r}^\prime) =
\delta(\textbf{r}-\textbf{r}^\prime)\hat{\textbf{e}}_j .
\label{eq:HGEOM}
\end{equation}
Now, all that remains is to map \eqref{HGEOM} onto an equivalent
real-frequency ($\xi$) system that can be evaluated in the time
domain, similar to \secref{time-domain}, for $\omega(\xi)$ given by
\eqref{Contour}.  There are at least two ways to accomplish this.  One
possibility, which we have adopted in this paper, is to define an
effective magnetic permeability $\mu_c = \mu \omega^2(\xi)/\xi^2$,
corresponding to a \emph{magnetic} conductivity, similar to
\eqref{dispersion}.  Combined with \eqref{Contour}, this directly
yields a magnetic conductivity as in \eqref{FDTD1d}.

A second possibility is to divide both sides of \eqref{HGEOM} by
$\omega^2/\xi^2 = 1 + i\sigma/\xi$, and absorb the $1 + i\sigma/\xi$
factor into $\varepsilon$ via \eqref{dispersion}.  That is, one can
compute the magnetic correlation function via the magnetic field in
response to a magnetic current with an \emph{electric} conductivity.
However, the magnetic current in this case has a frequency response
that is divided by $1 + i\sigma/\xi$, which is simply a rescaling of
$\mathcal{J}(\xi)$ in \eqref{Gdual}. There is no particular
computational advantage to this alternative, but for an experimental
realization~\cite{RodriguezMc09:PRL}, an electric conductivity is
considerably more attractive. [Note that rescaling $\mathcal{J}(\xi)$
  by $1+i\sigma/\omega$ will yield a new $g(\xi)$ in \eqref{gw},
  corresponding to a new $g(t)$ that exhibits slower decay.]

\subsection{Material Dispersion}

In this section, we extend the time-domain formalism presented above
to cases where the dielectric permittivity of the medium of interest
is dispersive. To begin with, note that in this case the dissipative,
complex dielectric $\varepsilon_c$ of \eqref{eps_c} is given by:
\begin{equation}
  \varepsilon_c(\vec{r},\xi) = \frac{\omega^2(\xi)}{\xi^2}
  \varepsilon(\vec{r},\omega(\xi)),
\label{eq:eps_c}
\end{equation}
where $\varepsilon(\vec{r},\omega(\xi))$ denotes the permittivity of
the geometry of interest evaluated over the complex contour
$\omega(\xi)$. 

This complex dielectric manifests itself as a convolution in the
time-domain equations of motion, i.e. in general, $\vec{D}(t) = \int
dt' \varepsilon_c(t-t') \vec{E}(t')$. The standard way to implement
this in FDTD is to employ an auxiliary equation of motion for the
polarization~\cite{YoungNe01}. For the particular contour chosen in
this paper [\eqref{Contour}], the conductivity term already includes
the prefactor $\omega^2/\xi^2$ and therefore one need only add the
dispersion due to $\varepsilon(\vec{r},\omega(\xi))$.

The only other modification to the method comes from the dependence of
$\Gamma^E(\xi)$ in \eqref{Gamma1} on $\varepsilon$. We remind the
reader that our definition of $\Gamma$ was motivated by our desire to
interpret \eqref{Force_fields} as the Fourier transform of the
convolution of two quantities, and thus to express the Casimir force
directly in terms of the electric and magentic fields $\vec{E}(t)$ and
$\vec{H}(t)$, respectively. A straightforward generalization of
\eqref{Gamma1} to dispersive media entails setting
$\varepsilon(\vec{r}) \to \varepsilon(\vec{r},\omega)$. However, in
this case, the Fourier transform of \eqref{Gamma1} would be given by a
convolution of $\vec{E}(\xi)$ and $\varepsilon(\vec{r} \in
S,\omega(\xi))$ in the time domain, making it impossible to obtain
$\Gamma^E(t)$ \emph{directly} in terms of $\vec{E}(t)$. This is not a
problem however, because the stress tensor \emph{must} be evaluated
over a surface $S$ that lies entirely within a uniform medium
(otherwise, $S$ would cross a boundary and interpreting the result as
a force on particular objects inside $S$ would be problematic). The
dielectric appearing in \eqref{Gamma1} is then at most a function of
$\omega(\xi)$, i.e. $\varepsilon(\vec{r} \in S, \omega) =
\varepsilon(\omega)$, which implies that we can simply absorb this
factor into $g(\xi)$, modifying the numerical integral of
\eqref{gndt}. Furthermore, the most common case considered in
Casimir-force calculations is one in which the stress tensor is
evaluated in vacuum, i.e. $\varepsilon(\vec{r} \in S, \omega) = 1$,
and thus dispersion does not modify $g(\xi)$ at all.



\end{document}